%% file: fauna_eb.tex
\documentclass{llncs}

\usepackage{makeidx}  
\usepackage[T1]{fontenc}   
\usepackage{amsmath}
\usepackage{amssymb}
\usepackage[dvips]{epsfig}
\usepackage{ae}
\usepackage{alltt}

\usepackage{pstricks}
\usepackage{fancybox}
\usepackage{pdfpages}
\def\repLATEX{.}

\usepackage{xspace}
\usepackage{boxedminipage}
\usepackage{\repLATEX/zedbis}

\begin{document}
\frontmatter          
\pagestyle{headings}  
\addtocmark{...ddtocmark...} 
%
%

\mainmatter              

\title{Dynamic Composition of Evolving Process Types}

\titlerunning{Dynamic Composition}
%
\author{Christian Attiogb\'e}
%
%
\tocauthor{Attiogbe (Lina)}
\institute{LINA CNRS UMR 6241 - University of Nantes\\
F-44322 Nantes Cedex, France\\
\email{Christian.Attiogbe@univ-nantes.fr}}

\maketitle              

\brokenpenalty=5000

\begin{abstract}

Classical approaches like process algebras or labelled transition systems deal with static composition to model non-trivial concurrent or distributed systems;   this is not sufficient for systems with dynamic architecture and with variable number of components. 
We introduce a method to guide the modelling and the dynamic composition of processes to build large distributed systems with dynamic adhoc architecture.
The modelling and the composition are based on an event-based approach that favour the decoupling of the system components. The composition uses the sharing of abstract communication channels. The method is appropriate to deal with evolving processes (with mobility, mutation). The event-B method is used for practical support.
A fauna and its evolution are considered as a working system; this system presents some specificities, its behaviour is not foreseeable, it has an adhoc (not statically fixed) architecture.

\end{abstract}
\keywordname{~Dynamic Composition, Modelling Method, Dynamic Architecture, Event-B, Rodin}

\centerline{\textbf{Preliminary version, 2011}}
\section{Introduction}
\label{section:intro}
\input{intro}

\section{Modelling Evolving Process Types}
\label{section:pbmStatement}
\input{pbmStatement}

In the next section, we introduce the requirements of a system that involves several interesting facets which can be handled by an appropriate method. 
\section{The Illustrative Requirements: the Fauna System}
\label{section:faunaSystem}
\input{faunaSystem}

In the following we propose a specification and verification method appropriate to dynamic composition and formal reasoning; it is based on event-driven modelling at abstract level. 
\section{The Proposed Method}
\label{section:proposedMethod}
\input{theMethod}

\section{Putting into Practice}
\label{section:puttingPractice}
\input{practiceFauna}

\section{Conclusion}
\label{section:conclu}
\input{conclusion}

\bibliographystyle{alpha}

\input{fauna_eb.bbl}
\appendix
\includepdf[pages=-]{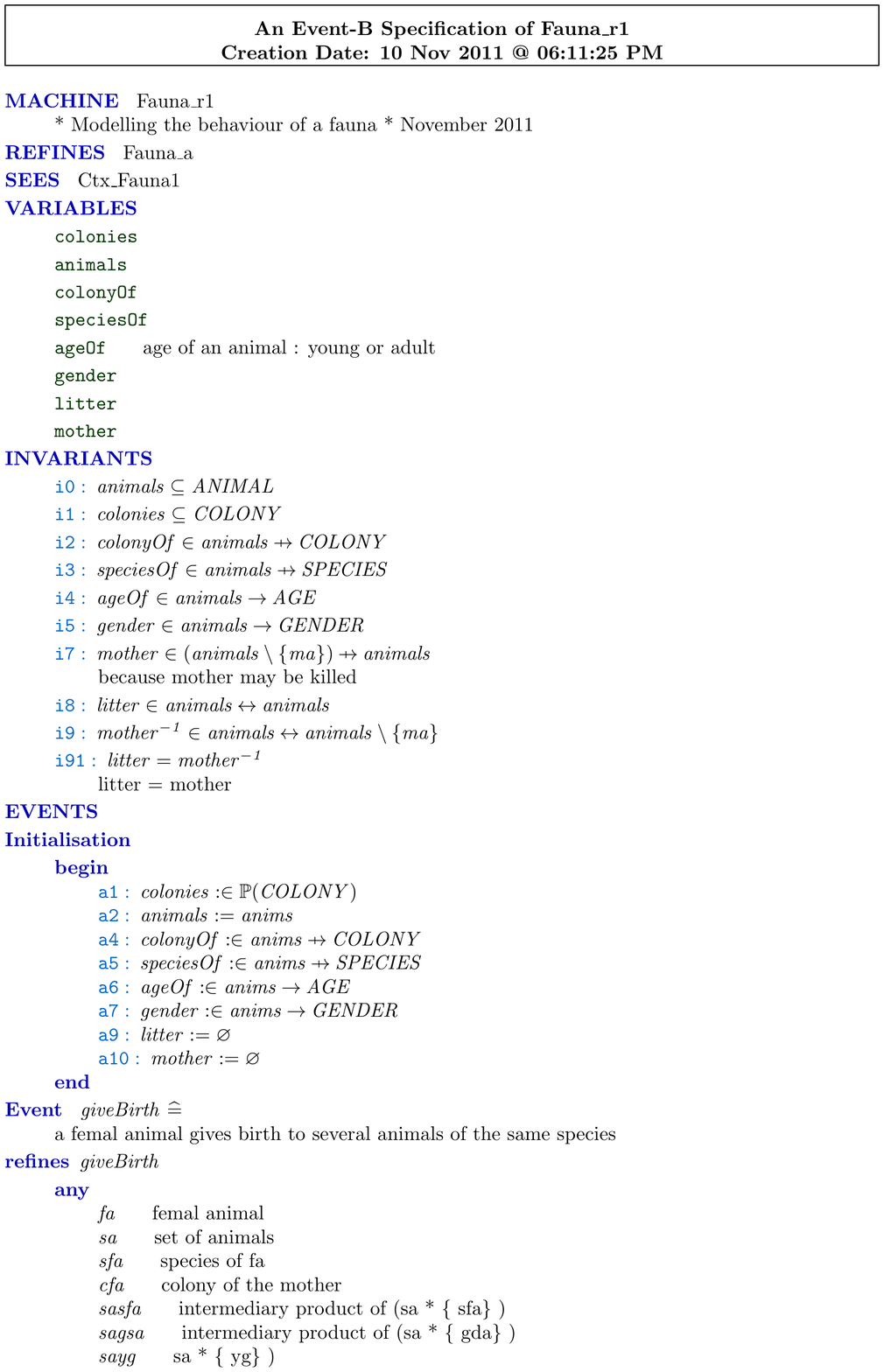}
\end{document}

%% file: intro.tex


Modelling and studying unstructured software systems made of (sub)components or processes with evolving behaviour is a challenging concern.

\textit{Context.}
We consider unstructured software systems  that do not have statically defined architecture  but an adhoc one which is continuously restructured on the fly due to the environment. Such systems adopt various different behaviours and are changing according to number of internal and external events. 
The evolution of these systems is consequence of internal actions and leads to noticeable  behaviour; the  internal actions may be or not the consequence of external events perceived by the system.

As regards behaviour, a system made of several interacting components, may react to its  environment and completely change its behaviour  accordingly: the component types evolve since they adopt new behaviour. 
In this context, modelling and correctness  analysis are very challenging. Indeed, most of modelling techniques are dedicated to well-structured systems. But a lot of computer systems obtain the adhoc features described above. 
There are many examples: an auction system, a multi-actors  conference on-line, an industrial process (sometimes reactive or real-time) which reacts with its environment and adapts an appropriate behaviour; an information system of a company, where there are departments, people, relations between people, and behavioural rules in the company; some peer-to-peer systems like gnutela; social networks where people join with a profile, change their profile, interact without severe restriction, leave the network; cooperating web-services on  the net or within a grid, etc. 
Many software systems range in this category, among which some critical ones. Characteristic properties are that their structure is unknown, their components are only partially known; they are globally discrete and asynchronous. Hence the needs of efficient modelling and reasoning techniques that guarantee the system correctness.
The study of behaviour of systems (not yet developed) can be done by simulation, prototype or by reasoning on a formal model if one wants to conduct a rigorous analysis. However, one must take care of the evolution aspect during the modelling of systems in order to obtain satisfactory conditions for their study.

We focus this work on the modelling with event-based approaches; they have the advantage to decouple the system components. Event-based components are designed to work with any components instead of specific components. Consequently they carry the potential to ease integration of autonomous, heterogeneous components into more larger or complex systems. Hence these approaches facilitate the study of unstructured software systems, their evolution and their scalability.

\textit{Contribution.} 
We  improve and extend previous results: *P-B~\cite{coloss:CA_ISOLA2008,attiogbePRDC09} to take account of process types, process generation, process types composition and evolution.
This results in a method that extends event-based approaches to face the modelling of unstructured and dynamically evolving systems. We propose an implicit composition of process types with an event-based approach; we show how the formal analysis can be conducted using the Event B framework. 

The interest of the current study is manifold: 
the proposed method is easily workable within existing event-based frameworks such as the Event B one;
the genericity of the treated case (a fauna system) and its adaptation for modelling and simulating  numerous real life situations.  

\textit{Structure of the article.}
The Section \ref{section:pbmStatement} discusses the issues on modelling evolving process types.
Section \ref{section:faunaSystem} presents a representative working case: a fauna system; the requirements are analysed and the challenging features are emphasized.
Section \ref{section:proposedMethod} is devoted to the proposed modelling method.
In Section \ref{section:puttingPractice} we report on the application of the proposed method to the fauna case study.
Section \ref{section:conclu} concludes the article.

%% file: pbmStatement.tex


Consider modelling and reasoning about a system made of several interacting processes of various natures with specific behaviours; the processes may for example change their behaviours during the time or according to specific events from their environment: they change their types which define their behaviours. The nature of the system may also change due to the fact that new processes may join it or existing processes may leave it at any time, the processes may interact with arbitrarily established links. Such a model of system is representative of internet-based systems, service-based systems, grid-computing systems, etc. 
Some features of this kind of systems and models are: the undefined or evolving architecture of the global system, the variable number of interacting processes, $\cdots$.
But, safety and reliability remain desirable properties for these complex systems, critical or not. Consequently, modelling and reasonning with precise methods and tools are needed.

\subsection{Limitations of Existing FSM-like Approaches}

The structure of a classical centralised software system is based on the composition of
several sub-systems or processes. They are often parallely composed to
enable synchronisation and communication. Classical approaches such as Finite State Machine or Process Algebra provides means to model and compose processes.


\noindent
\textit{State Transitions Approach.} Capturing  a process behaviour is intuitive 
but state transition systems lack high level structures for complex processes.
Handling an undefined and variable number of processes is not tractable.
Dealing with several instances of the same processes is not possible.
Synchronization of processes should be made explicit.

\noindent
\textit{Process Algebra.} Basic processes are described by their sequential behaviours; they are then parallely composed to form large processes.  Abstract channels or ports common to the processes are used as communication means. Very often, the parallel composition involves two processes (Hoare's CSP style); but is generalisable to several known processes (like with LOTOS).  

\medskip
In the previous approaches, the parallel composition operators  enable interaction between processes, they structure already defined processes (or process types).  
The composition of processes looks like 
 $P_1||P_2||\cdots||P_n$ where the description of $P_i$ (including their subprocesses) are already defined and $||$ denotes one parallel composition operator considering the common action alphabet used by the processes.
However it is not possible to integrate in the composition a completly new process which is not defined before (even as a subprocess) to model for instance a process joining a meeting of processes already started.

\subsection{Event-based Approaches}

Event-based programming (or implicit invocation) has been intensively used in programming applications with Java, C++, etc. It is for instance used for programming applications with GUI, for Active Databases management.
 
However there are lacks concerning the system modelling and verification approaches. As far as programming is concerned, Harel's statecharts for example contribute a lot to improve FSM-based programming by introducing structure and event-handling at program architecture level; therefore a global program structure can  be generated from the statechart models. 

As regards modelling, one can refer to modelling event-based treatment of discrete systems and distributed even-based systems treatment.

Discrete Event-based Systems \cite{CassandrasLafortune99} have a discrete set of state space and their transition system is based on event occurrence that change or not the current state. Automata are the basic model to model and study discrete event-based systems; they are amenable to composition operations and to analysis.
However they lack of structuring mechanisms in the case of large-scale system modelling which leads to state explosion; also with regards to analysis, only finite state space is tractable.   The Event-B method \cite{EVentB_Abrial2010} is a formal method that supports an event-based approach, refinements and theorem-proving to ensure correction-by-construction. 

As far as Distributed Event-Based Systems \cite{MuhlBook2006} are concerned, various components communicate by generating and receiving event notifications.  An event is any occurrence of a state change in some component. The affected component issues a notification describing the observed event.
Event notification system uses specific explicit operators or primitives: \textsf{sub, pub, notify,...}. Therefore a specific \textit{middleware} above operating systems facilitates the communications between the entities involved in an heterogeneous distributed computing environment.

To tackle modelling diffulties and reasoning issues, we need methods and verification facilities: not only model-checking or testing techniques which are pragmatic and more and more used, but also theorem-proving techniques and tools which can deal with unstructured systems, scalability and even infinite systems.

%% file: faunaSystem.tex

We consider a simplified version of a fauna and its evolution through the time; it is viewed as a system with several processes of various types and behaviours, without a predefined architecture.
The fauna is made of carnivorous mammals (lions, guepards, $\ldots$) and herbivorous mammals (gazelle, zebra, $\ldots$). 
Each animal has  precise invariant characteristics: gender, specy name, nutrition, $\ldots$.

The behaviour of an animal varies according to the gender and the age in the following way:
\begin{itemize}
\item Each animal always belongs to a colony consisting of animals of the same species, of both genders (male and female), young and adult.

\item A young animal belongs to its mother's colony. A single (unmarried) adult male can either leave its colony and enter another colony or can start its own colony.

\item A male adult can have one or several female partners of the same species, but not member of its colony. We can consider the case where the partnership is permanent until the death of the animals.

\item A female with a male partner can sire a litter. The female can only feed one litter at a time. The female abandons the litter when the young animals become adult.

\item Every carnivorous can only kill animals of another species. 
\item A herbivorous does not kill.

\item Every animal can die naturally or can be killed by a carnivorous.

\item Every litter that loses its mother disappears with it.
\end{itemize}
 
We assume that there is only one fauna consisting of  animals of several different species. We don't deal with  contacts with others fauna.

\subsection{Requirement Analysis }
The model of the system should take account of the following identified requirements: \\
\begin{center}
\begin{tabular}{|l|p{9cm}|}
\hline
F-REQ-AnType & There are several differents kinds of animals\\
\hline
F-REQ-AnBeh ~ & Each kind of animal has a general behaviour and a specific behaviour related to its actual features\\
\hline
F-REQ-Evol~ & Within a type, animals born, grows and die\\
\hline
F-REQ-inter ~ & There are interactions between animals\\
\hline
NF-REQ-struct~ & There is no precise structure of the global system \\
\hline
\end{tabular}
\end{center}

Animals are considered as processes with specific types.
The fauna is viewed as an unstructured decentralised system; it has a dynamically evolving architecture. 
Decentralised systems cannot be structured with parallel operators that compose a fixed number of processes; they have an ad-hoc structure related to the number of involved processes. 
The structure of the group of animals, hence the architecture of the system, is varying; processes may join or leave the group at any time. 
At least a group communication mechanism is needed to model such a system. 


\subsection{Challenging Features}
\textit{Structuring} and \textit{interaction} are two main features of systems with dynamic architecture.
We need a systematic method to deal with the case; such a method should serve as a guide to start with the statement of the requirements and reach an abstract model of it. 

With respect to modelling, the behaviour of each process should take account of the invariant properties of the processes and also of the unstability of this behaviour (due to type mutation).

Composing an unknown number of processes with mutant behaviours is a challenge. As composition operators are often defined with a fixed number (2) of processes, or a fixed (N) for some specification languages, it is not possible to deal with a dynamically varying number of processes with these languages.
Standard composition solutions are based either on Finite State Machines (FSM) with the product of FSM to deal with parallel composition or Labelletd Transition Systems (LTS), or on process algebra with binary parallel composition operator, synchronous communication (like in CCS) or n-ary communication (as in CSP or LOTOS).

All the same, a \textit{dynamic composition} is more appropriate;
the managment of communications between dynamically composed processes requires a careful treatment.

%% file: theMethod.tex
In this work we improve and extend previous results on the *P-B method~\cite{coloss:CA_ISOLA2008,attiogbePRDC09} by a systematic and rigorous treatment of process types and their composition via events and abstract communication channels. The leading principle is the description of several \textit{process types} that may be involved in an interaction. An abstract model that carries the possible interactions among the eventual processes of the described types is specified. 

\subsection{Overview of the *P-B Method }
The steps of the method are as follows :
\begin{enumerate} 
\item To build a \textit{reference formal model} from the system at hand and to state the desired global properties according to this formal model. The reference model is an abstract, multi-process model from which specific models may be built; it may be composed of elementary models. This step is detailed in Sect. \ref{section:buildModel}.
\item To perform formal analysis (property verification) with the reference model,
\item To refine gradually, if necessary, the abstract formal model into less abstract ones, and to perform (iteratively) formal analysis on the current model.
\item To deal with multifacet aspect if concerned; 
\begin{enumerate}
\item To systematically derive (or translate) from one level of the reference model, if necessary, other formal models which are specific inputs to various analysis techniques and tools; 
\item To perform formal analysis with the specific
      models or with their extensions, by adding specific properties to
      the global ones;
\item To ensure  the consistency between the reference model and the
      specific ones by propagating the feedback from the specific
      models study on the reference model and by updating consequently
      the other specific models.  Then, the analysis of each facet via a
      specific model participates in the global system analysis.
\end{enumerate}
\end{enumerate}

In the current article we do not deal with multifacet analysis (Step. 4), but we detail and enhance the specification approach to deal
with the  Step 1. (and partially with Steps 2. and 3.) of the method: building a reference model. This step needs methods that are appropriate to the system at hand. We guide  the construction of a system as \textit{virtual component net}, there are abstract components and abstrat channels wich lnk in a decoupled manner the components. Here component has its general sense as a part of an entity.

\subsection{The Used Methods and Tools}
We target the B method and its tools as an experimental framework. 
\input{ovv_eventB.tex}

\subsection{Event-based Modelling of Multi-Process Dynamic System}
\label{section:buildModel}

\begin{figure}[!ht]
\begin{center}
\includegraphics[width=0.5\linewidth]{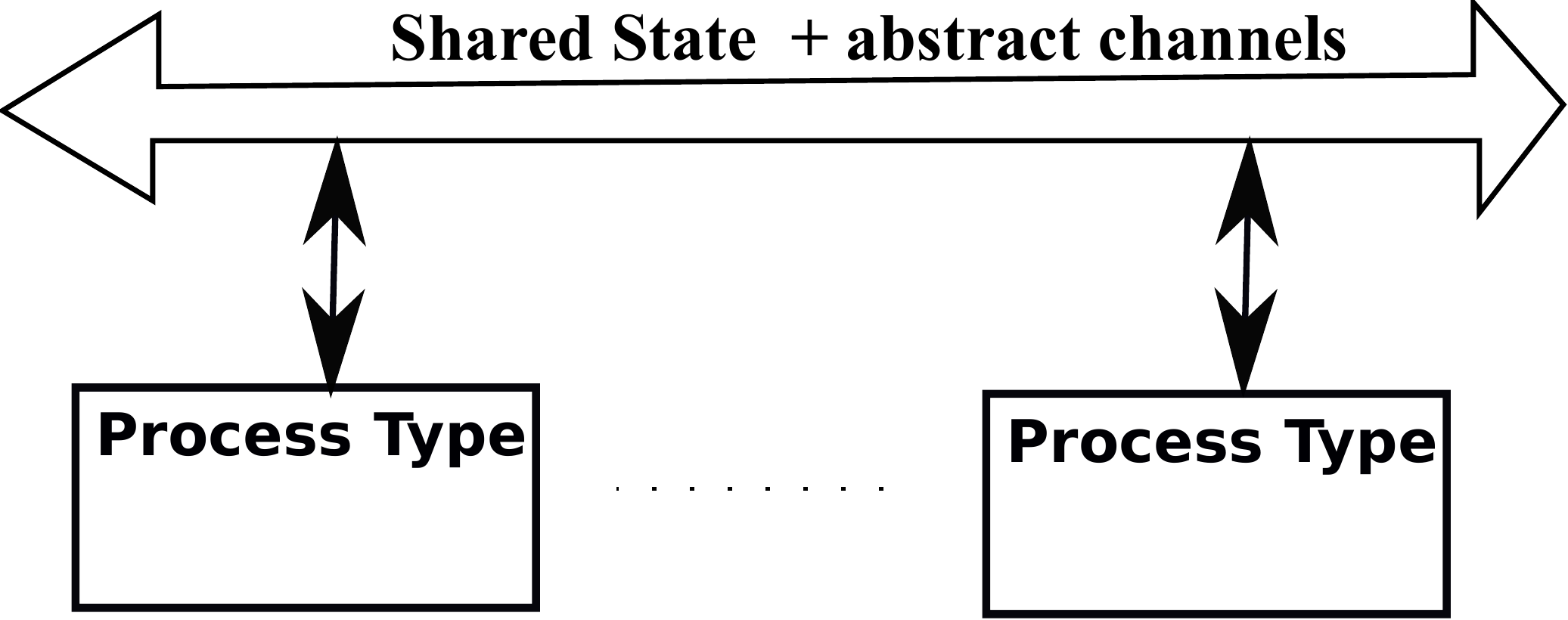}
 \caption{The schema of process type composition}
\label{fig:vchannel}
\end{center}
\end{figure}

We are going to define and link process types via identified abstract channels.
The process types are those identified within the requirements. The abstract channels are modelled according to the interaction needs. Therefore each process type use independently from the other the defined abstract channels and state. This is the cornerstone of the dynamic composition to get an adhoc architecture (see Figure \ref{fig:vchannel}).
 
Our approach then provides guidelines to help in discovering and modelling the desired behaviours of a system with dynamic architecture. It  emphasizes an event-based view at global level, for composing processes.
Within the Event B approach, used for practical aspects, process types are modelled as Event-B machines,
asynchronous systems communication is modelled with the interleaved composition of
process behaviours viewed as event occurrences. 
 
\begin{figure}[!ht]
\begin{center}
\includegraphics[width=0.6\linewidth]{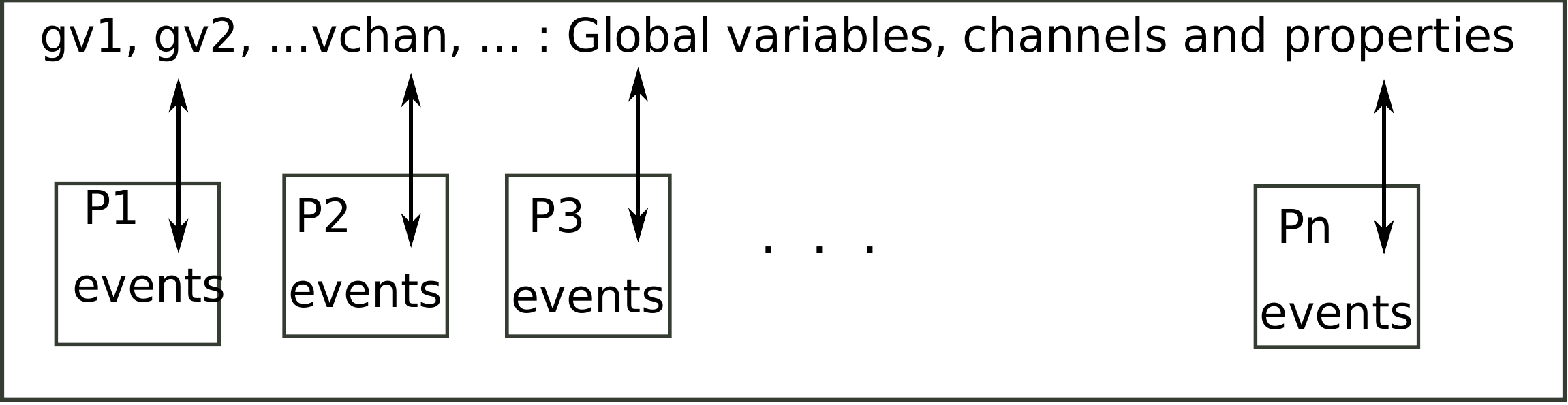}
 \caption{A scheme of composed process types}
\label{schemeCompo}
\end{center}
\end{figure}

The method used to build the reference model is summarised as follows.
\begin{enumerate}
\item \textbf{Structuring aspects.}\\
From the requirements, common state variables and properties are identified and described, they will be shared; elementary processes are identified according to the described behaviours. Basically they are \textit{types of process} since, depending on the cases, one can have several processes with the same behaviour hence the same type.

Each identified \textit{type of process} $P_i$ that may participate in
      the global system model is specified by considering its space state $S_i$
      and the events $E_i$ with their descriptions $Evt_i$ that carry the
      behaviour of the process type.  The constituents $S_i$, $E_i$ and $Evt_i$ will be detailed latter on. The modelling of the behaviour is detailed in Sect. \ref{subsubsection:buildBehav}. To handle the dynamic architecture of global system, we require that for each type of process, the events to join and leave the system be defined. Some events may be common to several proces types; they handle
      interaction and state sharing aspects. After this step, we get several process types ($P_i$):\\
\centerline{$P_i \defs \langle S_i, E_i, Evt_i\rangle$}
At this low
      level, either an event-oriented or a process-oriented view may be considered to discover the needed events for a process behaviour.
\item \textbf{Interaction aspects.}\\
Common abstract channels are to be introduced to link interacting processes.
Interaction involves communication. 
Communications are modelled with abstract channels.
An abstract channel is modelled as a set; it is used to wait for a message
      or to deposit it. Hence the interaction between the processes is
      handled using these common abstract channels. Therefore, the communications are
      achieved in a completely decoupled way to favour dynamic structuring. A process may deposit a message in the channel, other processes may retrieve the message from the channel.

Therefore we use guarded events, 
      message passing and the ordering of event occurrences; the processes synchronise and communicate through
      the enabling/disabling of their events. Therefore, an event
      is used to model the wait for a data by a process; hence it may be blocked until the availability of the
      data (enabling the event guard), which is the effect produced by
      another process event. Consider for
      example the case of processes exchanging messages, one process
      waits for the message, hence there should be an event with a disabled guard, and an other process with an enabled event which effects sends the message. \\
\item \textbf{Composition of the processes.}\\
Practically the composition is implicit during the modelling of the unstructured systems considered here.  
But a bottum-up view may be adopted where the composition of process types is explicit.

The described processes are combined by a \textit{fusion} operation $\biguplus$ that merges an undefined number of process types.
  The fusion operator merges the  state spaces and the events of the processes into a single global system $Sys_g$ which  in turn can also be involved  in other fusion operations.
$$ Sys_g ~\defs~~ \biguplus_{i} P_i~~=~~\biguplus_{i}~\langle S_i, E_i, Evt_i \rangle ~~=~~\langle S_g, E_g, Evt_g \rangle$$ 

According to the fusion operator, when  process types are merged, one set is introduced for each type to identify the processes of this type. Each feature that is modelled with a variable in a process type, results in a function from the set of process identifiers to the domain of  feature's values. 
The processes access the global state and communicate with others, through their events (as depicted in Fig. \ref{schemeCompo}).
 
\end{enumerate}

In the following we make it precise the modelling of process types through their states, invariant and behaviour. Then we deal with the dynamic composition of process types through the composition of invariants and events.   
We illustrate  the method  with the FAUNA system using the B abstract system as a support for the description of process types.

\subsection{Modelling the State Space}
\label{subsubsection:modlstate}
Abstract channels shloud be typed as the processes echange structrured data and messages. The types of the channels depends on the global data types used in the abstract state description. 
Two main points are to be underlined: to generalise the description of a process to the description of a type of processes that have the same behaviour; to capture the features and properties that are common to several processes.

The state space ($S_i$) of each process type $PT$ is such that, a fresh variable \textit{thePTs} is introduced to denote a set of all the processes of the same type PT. Each element of this set identifies exactly one process.

\textbf{If} $\langle f_1 : Type_{f_1},~ f_2: Type_{f_2},~\cdots,~f_n : Type_{f_n} \rangle$ is the state characterisation of a given process $P$,\\
\indent \textbf{then}  $\langle f_1 : thePTs \fun Type_{f_1},~ f_2:  thePTs \fun Type_{f_2},~  \cdots,~f_n :  thePTs \fun Type_{f_n} \rangle$ characterises the type $PT$ of the processes P.\\


Specific features defined for subprocesses are managed by defining these features according to the appropriate subsets of the set of processes. 
Shared features are  described in a standard way using constants typed variables.

\subsection{Building the Behaviour: Types, Properties, Evolution}
\label{subsubsection:buildBehav}

\begin{figure}[!ht]
\begin{center}
\includegraphics[width=0.9\linewidth]{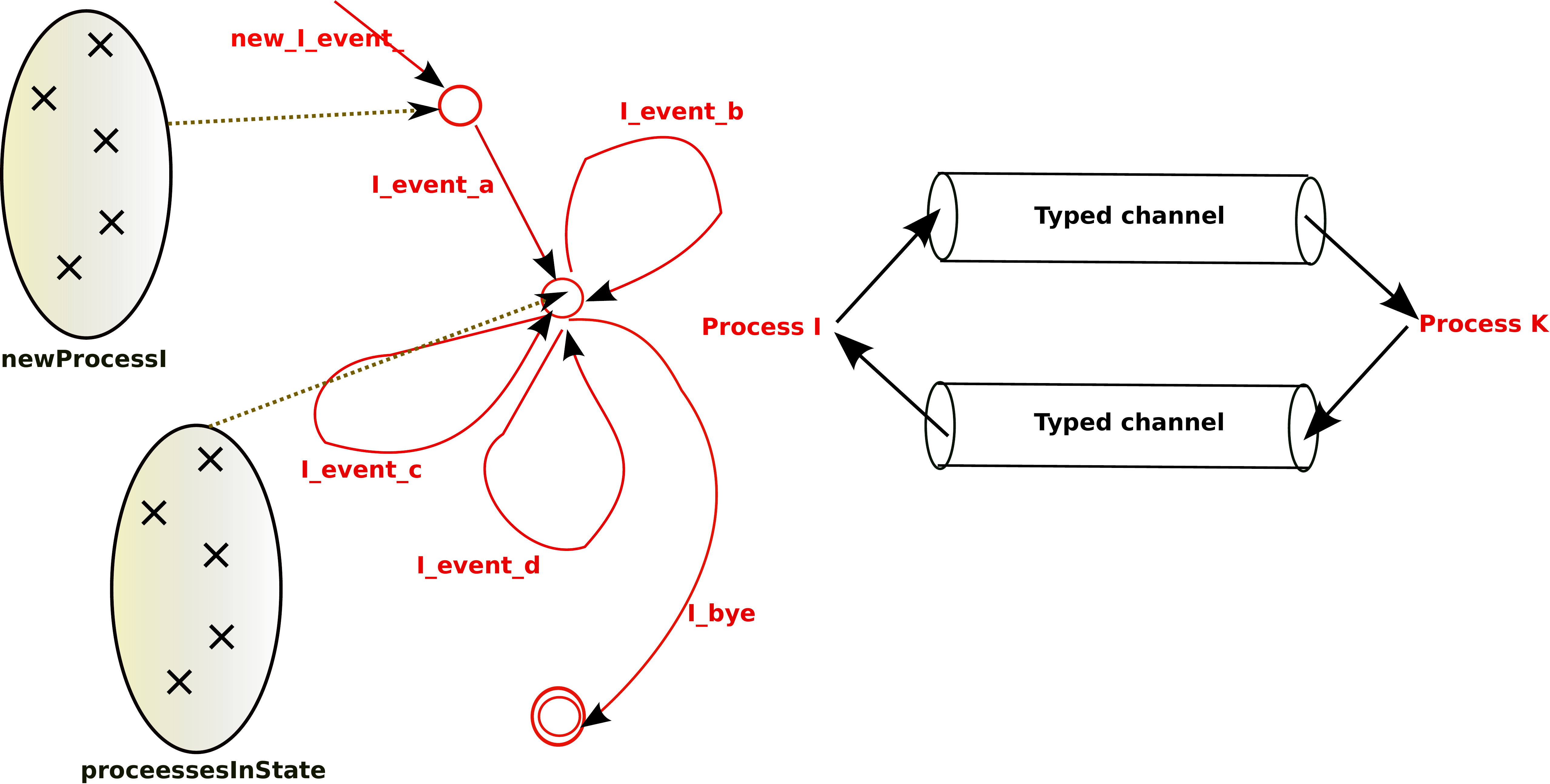}
 \caption{Structuring the behaviour of a process type}
\label{fig:extendedMealy}
\end{center}
\end{figure}

As regards the events ($E_i$), assume that we have defined a set of variables ${\mathcal V}$ for the state space $S_i$, it remains to define a set of events whose actions are guarded by predicates expressed using ${\mathcal V}$. This can be achieved gradually, and by merging event descriptions. But in order to deal with dynamic architecture and dynamic evolution of the composed processes, the events have to be defined in such a way that several processes (instances of the same type) run the same events as their behaviour.
For this purpose we recommand to use of a Mealy machine (LTS with guard) to structure the behaviour of each process type. This is necessary to properly capture the desired behaviour as it is tedious to only think of events and their relationships without a concrete support (as is the Mealy machine). To help in modelling the event so as to deal with several process occurences, we extend the Mealy machine (see Fig. \ref{fig:extendedMealy}), by annotating its state with the set of processes in this state; thus the transition from a state is nondeterministically achieved by one of the process in this state. The guards of events capture this semantics. The events to model are described by translating the extended Mealy machine into a set of events, each event translate a transition \footnote{this will be done by a tool in future work}.  

Therefore a set of events is associated to each identified process type and its subtypes. The behaviour of a process type is denoted by the combined occurrences of its events.
It remains to describe the events ($Evt_i$) of each process type to have its behaviour. We introduce for this purpose a scheme for the event description, where the guards of events are specific to the process types. 

The specification of an event has the generic form $\textit{evt}~\defs~ guard~ \rightarrow~ action$. 
The more readable syntactic form of event specification is:

\vspace{-0.5cm}
\begin{tabbing}
\=\hspace{1.4cm}\=\hspace{1cm}\=\hspace{1cm}\=\hspace{1cm}\\
\> \textit{evt } $\defs$ \> \textsf{when} \textit{guard} /* the condition to enable the event */\\
\>\> \textsf{then} \textit{effect description} /* the action of the event */\\
\>\>  \textsf{end}
\end{tabbing}

If a list of events $e_{P_j}$ characterises a process type $PT$, then all the events $e_{P_j}$ are guarded by the type $PT$\footnote{as a type is a property, it is trivial to write a guard $p \in PT$ }. That means these events will be enabled only for process of type $PT$.

The effect of an event may impact  the state of the process considered or the global system, via the modification of state variables.
The following event description schema handles subtypes and specific properties for the  instances of a given type. 

\vspace{-0.6cm}
\begin{tabbing}
\=\hspace{1.4cm}\=\hspace{1cm}\=\hspace{1cm}\=\hspace{1cm}\\
\> \textit{evt } $\defs$ \> \textsf{when}  \textit{process~of~type~}$PT$ ~$\&$~\textit{specific predicates}  \\ 
\>\>  /* for every process of the type, with specific condition */\\
\>\> \textsf{then}  \textit{effect description}\\
\>\>  \textsf{end}
\end{tabbing}

An event without a guard describes a piece of behaviour shared by all processes.

\medskip
\noindent \textbf{Types are properties (predicates).}
Every type can be expressed as a predicate that denotes a property; the guards of events are then the predicates describing the considered types and the needed enabling conditions.
Therefore the description of processes using state variables and invariant, with additional conditions or properties is sufficient to express the behaviour as guarded events.

As several processes of the same type share the same behaviour,  the descriptions of the behaviour's events are non-deterministic; this is captured using an appropriate B construct to express that \textit{any instance $p$ of the type for which the guard is true may evolve}. 

\subsection*{The Resulting Modelling Scheme of a Process Type}
Consider a process type $PT$ where each process has the features $a$,$f$ and a behaviour defined by the combined occurrences of the events $e_l$, $e_g$.The modelling scheme of $PT$ is as follows: \\
 
\begin{boxedminipage}{8cm}
\begin{tabbing}
\hspace{1cm}\=\hspace{0.8cm}\=\hspace{1cm}\=\hspace{1cm}\=\hspace{1cm}\\
\textsc{used sets}\\
\>   $PT,~ Da,~ Df, GT$  ~~~~/* Da (Df) is the value domain of a (f) */\\
\>  $\cdots$\\
\textsc{invariant properties}\\
\> $thePTs~ \subseteq~ PT$ ~~~~~~/* the set of processes of type PT */\\
\> $a : thePTs \fun Da ~\land$ /* each process has an $a$ */ \\
\> $f : thePTs \fun Df  $ ~~~/* each process has an $f$ */\\
\> $gv : GT$ ~~~~/* a global variable with a given type */\\
\> $gChan : ChanType$ ~~~~/* a global channel with a given type */\\
\textsc{event descriptions}\\
\>  $e_l$ = \textsf{any} $p,l$    ~~~~~~~~/* event depending on local variables l */ \\
\>\>       \textsf{where} $p \in thePTs$ /* $p$ is one of the processes */\\
\>\>\>$\land~ cond_{(p,l)}$  ~~~/* with specific condition $cond_{(p,l)}$ */\\
 \>\>      \textsf{then} $\cdots$ \\
 \>\>      \textsf{end} ;\\
\>  $e_g$ = \textsf{any} $p$ ~~~~/* an event with gv a global variable */\\
\>\>       \textsf{where} $p \in thePTs ~\land~ cond_{(p,gv)}$  \\
 \>\>      \textsf{then} $\cdots$ ~~~~~ /* may impact on global state */\\
 \>\>      \textsf{end} 
\end{tabbing}
\end{boxedminipage}

\medskip
This modelling scheme (expressed here as a pseudo-B machine) is systematically used for each identified type. The non-deterministic style of the event descriptions allows both the co-evolution of any number of processes and their dynamic composition with processes of the same type or not.

\subsection{Dynamic Composition: Composing Invariants and Events }

As regards consistency and behaviour, the dynamic composition of process types is based on the composition of invariants and events; that is the conjunction of invariants and the merging of events. We make the assumption of shared variables. 
The formal basis of this approach is well studied \cite{ZaveJack93,tla+99,charpentier00theorems,Charpentier_CompoInv2006}. 
The dynamic composition of the processes is then the composition of their invariants and events to obtain a larger process which behaviour is the combination of the behaviours of the composed processes. The composition is hierarchical.
This composition approach decouples the processes, enables \textit{group communication, dynamicity, mutation, architectural evolution}. Indeed it is like if all the events of a process type are tagged with a color. One can remove all the events of the same color, replace them with other events of the same color or not; a set of events with new color can join the existing processes (changing thus the architecture), etc.  
The processes are to be defined incrementally. A process mat be replaced by another (replacing the set of events).

The composition as proposed has a decomposition counterpart; this is interesting for refinement purpose.
In the case of the event-B method there are various proposals for the decomposition of B abstract machines;   Abrial\cite{Abrial_EBDecomposition2009} proposed a decomposition based on shared variables; the composition of machines with shared variables are then the disjoint union of the variables and the events are merged.
Butler\cite{Butler_EBDecomposition2006} proposed a decomposition based on shared events where the viewpoint is a parallel composition of several Event-B machines into a composite machine; in this case the separate composed machines operate on disjoint variables and the machines interact by synchronising on events.\\

\subsection{Interaction Means}
In our approach, interaction is supported by communication and synchronisation between processes and groups of processes. The events whose guards use shared variables (including channels) may act as communication and synchronisation means. 
A process of a group (involved in a composition) may send/receive messages  to/from other processes of the group.  
The interaction among the processes that compose a system is based on message passing through abstract channels. Abstract channels defined as sets are used for this purpose. Such sets are variables shared by the processes. It is worth to note that the number of events is quite infinite, the sets of processes involved in the event guards are varying, leading to the very simple interaction and a decoupled architecture of the communicating processes. For example if a new process $p'$ of type $PT$ joins the existing group, an event is observed and consequently the appropriate variable ($thePTs$) is updated; therefore the new process may participate in all the events whose guards contain $p \in thePTs$.  

Unlike process algebra where a fixed action alphabet is used as parameter to make the processes communicate, the described event-based approach makes it easy the communication between various processes.


%% file: ovv_eventB.tex
%
%
The B method \cite{Abr96a} is a state-based approach for modelling and constructing correct software. It uses set theory, logics and refinement. The method has been extended, known as Event-B, to deal with reactive, distributed or concurrent system \cite{EVentB_Abrial2010}.

Like a B \textit{abstract machine} dedicated to structure sequential systems, a B \textit{abstract system} (also called a B model) \cite{Abr96a,AbrialMussat98,EVentB_Abrial2010} describes a mathematical model of a system behaviour\footnote{A system behaviour is the set of its possible transitions from state to state beginning from an initial state.}. It is  mainly made of a state description (constants, variables and invariant) and several \textit{event} descriptions. 
The state of an abstract system is described by variables and constants linked by an invariant. Variables and constants represent the data space of the system being formalized. Abstract systems may be refined like abstract machines \cite{AbrialCansellMery03,EVentB_Abrial2010,DBLP:conf/icfem/SuAHZ11} in order to build and prove gradually the desired system. 

\noindent
\textbf{Data of an Abstract System} At a higher level an abstract system
models  an entire model, be it distributed or not.
The data space that are formalized within the abstract system may correspond to all the elements of the (distributed) system.
Abstract systems have been used to formalize the behaviour of various (including distributed) systems \cite{AbrialCansellMery03,TopologySCP_HoangKBA09,EVentB_Abrial2010}.

\noindent
\textbf{Events of an Abstract System} Within event-B, an event is considered  as the observation of a system transition. Events are  spontaneous and show the way a system evolves. An event has a \textit{guard}(a predicate) and an \textit{action}. The event may occur or may be observed only when  its guard is true (the event is enabled). 
The action of an event is described with substitutions; it models how the system state evolves when this event occurs.
Several events may be enabled simultaneously; in this case, only one of them  occurs. The system makes internally a nondeterministic choice. If no guard is true the abstract system is blocking (deadlock).

\noindent
\textbf{Semantics and Consistency.} An abstract system describes
a mathematical model that simulates the  behaviour of a system. It has a trace semantics which is strenghened by an invariant. The semantics is therefore  based on the invariant and is  guaranteed  by proof obligations (POs). The \textit{consistency} of the model is established by such proof obligations:  
\textit{i)} \textit{the initialisation should establish the invariant};  \textit{ii)} \textit{each event of the given abstract system should preserve the invariant of the model} (one must prove these POs). 

\noindent
\textbf{Refinement}
An event-B abstract system may be refined into more concrete ones; the state space is refines by considering less abstract mathematical structures. The behaviour may also refined with respect to the data space, the introduction of new events events to refine the previous ones. The refinement is constrained by proof obligations (invariant preservation, no contradiction with previous events) which should be discharged.
 
\noindent
\textbf{The Rodin Tool}
There are several tools dedicated to the B method; Event-B is supported by the \textsf{Rodin} tool\footnote{\url{http://rodin-b-sharp.sourceforge.net}} \cite{rodin_AbrialBHHMV10}. It is an eclipse-based tool which offers edition, modelling, simulation, model-checking and proving functionalities. The proof obligations for consistency and refinement proof are generated by the tool and assistance is provided to discharge the proof obligations.

%% file: practiceFauna.tex


\subsection{Event-B Modelling of the Fauna System}
We have to model the fauna system  as the composition of an arbitrary number of various process types that compose the fauna, provided that they may evolve and interact in various way.

Applying the presented method (Sect. \ref{section:buildModel}) 
we identify several process types with subtyping relationship.
Two main process types are considered in the modelling and the composition: Carnivorous and Herbivorous. Each one has a specific behaviour, and evolves as young and then as adult with appropriate behaviours. Interactions are possible between both for example carnivorous killing herbivorous.

The state space description is modelled as follows; a given set ANIMAL stands for the more general type of the animals. We use a set COLONY to identify the colonies in which animals live. \\
A function $colonyOf : animals \pfun ANIMAL$ gives the colony of each animal.

To partition the animal set in several identified types, we use functions from $ANIMAL$ to the following enumerated sets (the converse of the functions give the appropriate subsets of animals of each category):
$AGE ==\{young, adult\},\\
~ GENDER==\{male, female\},~ SPECIES==\{carnivorous, herbivorous\}$.
As each animal has exactly one value for each of these properties, we use the functions 
 $speciesOf$, $ageOf$, $gender$ defined from the set of animals to the appropriate sets. Combining predicates written with this function we are able to deal with the needed eight types to build the abstract model.



As regards the behaviour, every animal may born, live by playing, and may die; the events \textit{play, die} are associated to the more general type \textsc{animal}. Specific events are shown in the Table \ref{table:theTypesAndEvent}; \textit{leaveC} stands for "to leave a colony".  

\begin{table}
\begin{center}
\begin{tabular}{|c|c||c|c|} \hline
\multicolumn{4}{|c|}{\textsc{animal} } \\ 
\multicolumn{4}{|c|}{\textit{{\blue \{born, play, die\} }}} \\ \hline \hline
\multicolumn{2}{|c||}{\textsc{carnivorous} } & 
\multicolumn{2}{|c|}{\textsc{herbivorous}} \\ \hline

 \multicolumn{2}{|c||}{\textsc{young}} & 
 \multicolumn{2}{|c|}{\textsc{young}} \\
\hline
\textsc{~male~} & 
\textsc{~female~} & 
\textsc{~male~}  & 
\textsc{~female~}  \\
\hline
 \multicolumn{2}{|c||}{ $\downarrow$ {\red \{becomeAdult\}}} &
 \multicolumn{2}{|c|}{ $\downarrow$ {\red \{becomeAdult\}}} \\
\hline
\hline
\multicolumn{2}{|c||}{\textsc{adult}} & 
\multicolumn{2}{|c|}{\textsc{adult}}  \\ 
\multicolumn{2}{|c||}{ {\red \{kill\}}} & 
\multicolumn{2}{|c|}{ ~~ }  \\
\hline
\textsc{~male~} & 
\textsc{~female~} & 
\textsc{~male~} & 
\textsc{~female~} \\
\hline
  ~\textit{{\red \{leaveC, startC\} }}~ & 
  ~\textit{{\red \{sireLitter\}}}~  &
  ~\textit{{\red \{leaveC, startC\} }}~ & 
  ~\textit{{\red \{sireLitter\}}}~ \\  
\hline
\end{tabular}
\end{center}
\caption{The identified process types and specific events}
\label{table:theTypesAndEvent}
\end{table}

Eight  process types (\texttt{male\_young\_carnivorous, female\_young\_carnivorous, male\_adult\_carnivorous}, ...) are then identified; they correspond to the types $P_i$ described in the previous section. The fauna system model is the dynamic composition of the behaviour  related to these eight types.
The requirement F-REQ-AnType is achieved throught these different types.

$$ Fauna_g ~\defs~ \langle S_f, E_f, Evt_f \rangle $$
$$ =~~Carnivorous~ \biguplus~ Herbivourous$$
$$ =~ \biguplus_{i} P_i~~=~~\biguplus_{i}~\langle S_i, E_i, Evt_i \rangle$$ 
$$ =~~ CarnivorousYoungMales~ \biguplus~ CarnivorousYoungFemales$$
$$\biguplus~~~CarnivorousAdultMales~ \biguplus~CarnivorousAdultFemales$$
$$\biguplus~ HerbivorousYoungMales~\biguplus~ HerbivorousYoungFemales$$
$$\biguplus~~~HerbivorousAdultMales~ \biguplus~ HerbivorousAdultFemales$$\\

The specific events that describe the behaviour of the animals, instances of the eight types are then 
$$E_f = \biguplus_{i} E_i  = \{ born, play, die, leaveC, kill, becomeAdult, startC, sireLitter\}$$

The descriptions of the process types and the events of $Evt_f$ that define the type behaviours  are done using Event-B.  Each type is characterised by a set of events. \\
Note that the requirement F-REQ-Evol is achieved;
F-REQ-inter is achieved through the $kill$ event.

\begin{figure}[!h]
\includegraphics[width=1\linewidth]{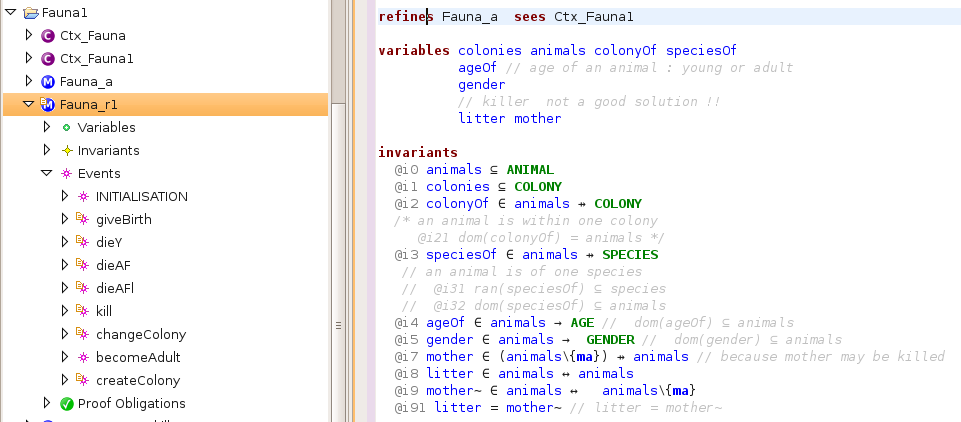}
\caption{The process type ANIMAL}
\label{fig:animalType}
\end{figure}

The figure Fig. \ref{fig:animalType} gives an overview of the Event-B abstract machine corresponding to the fauna system. A more detailed listing is given in appendix.

\begin{figure}[!h]
\includegraphics[width=0.8\linewidth]{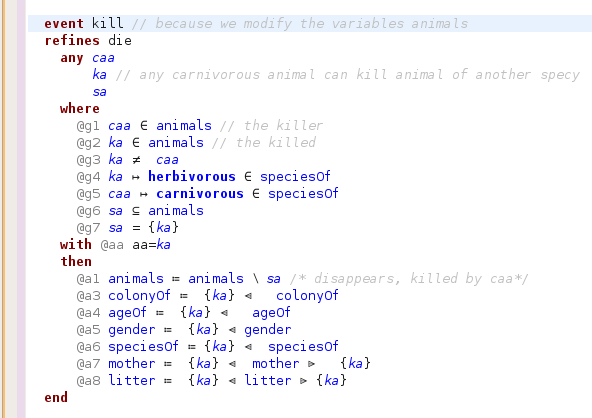}
\caption{The \textsf{kill} event - a refinement}
\label{fig:killEvent}
\end{figure}

The specification of the event \textsf{kill} is given in Fig. \ref{fig:killEvent}.

\subsection{Formal Analysis using Rodin}

For the global consistency of the abstract model the properties are described in the invariant. They are checked with respect of invariant preservation proof oblogations of the Event-B method.

Specific properties are added in the invariant to enhance the correctness the model. 

\begin{tabular}{|l|l|}
\hline
~PropKCarn~~ & The killers are carnivorous~~\\
\hline
 $~~$          & $killer \in animals \pfun animals$\\
               &  $\forall a . a \in dom(killer) \implies  a \mapsto carnivorous \in speciesOf$ \\

\hline
~PropLM~~ & The animal of the same litter have the same mother \\
\hline
          & $mother \in animals \fun animals$\\
          &  $litter \in animals \rel animals$\\
          &  $litter = mother \tilde~ $\\
\hline
~ConsistKill~& a killed animal is no more in a colony \\
\hline
             & $   dom(killer) \cap  dom(colonyOf) = \{\}$ \\
\hline
\end{tabular}

\medskip

\begin{figure}[!ht]
\begin{center}
\includegraphics[width=0.7\linewidth]{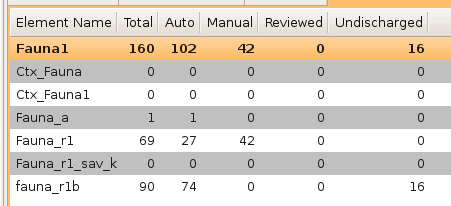}
\caption{Verification Statistics}
\label{tab:faunaStats}
\end{center}
\end{figure}

In our experimentation, we completly develop and prove correct the fauna system. 
The proof statistics (from the Rodin framework) are give in Fig. \ref{tab:faunaStats}. Several proofs are automaically discharged by the Rodin prover. The other are interactively achieved mainly by guiding the prover in chosing the proof steps. 

%% file: conclusion.tex
%

We have presented a method to dynamically compose processes modelled as Event B machines. The composition is based on an event-based communication through common abstract channels. The processes do not know each other but interact through the channels: the model describes a virtual component net.  This is suitable to deal with unstructured system with adhoc architecture. The method is based on well-researched theoretical basis and is supported by the Rodin  framework dedicated to Event B method. We have given an illustration of the method on an example of a fauna system that has the feature of an unstructured, distributed and evolving system. 

Given a requirement of a concurrent or distributed system, the proposed method helps to start its analysis, to follow precise steps until reaching a formal abstract model which constitutes the basis of further development.  

The development of plug-ins for Rodin to deal with the machine merging is planned as further works. The idea is to make it easy the management of the global model by keeping separated the machines that represent the identified types. That is suitable for dealing with large models; this approach should also be studied for consistent separated refinement.